\documentclass[preprint,aps,onecolumn,floats,floatfix,amsmath,nofootinbib]{revtex4-1}
\usepackage{graphicx,array,dcolumn}
\usepackage{calc,tabularx, epsfig,mathrsfs}
\usepackage{hyperref}

\usepackage{amsmath,verbatim,enumerate}
\usepackage{amssymb}
\usepackage{multirow}
\usepackage{xspace}
\usepackage{slashed}
\allowdisplaybreaks[1]
\newcommand{\beq}{\begin{equation}}
\newcommand{\eeq}{\end{equation}}
\newcommand{\bea}{\begin{eqnarray}}
\newcommand{\eea}{\end{eqnarray}}
\newcommand{\ba}{\begin{array}}
\newcommand{\ea}{\end{array}}
\newcommand{\bg}{\bar{g}}
\newcommand{\mn}{{\mu\nu}}

\newcommand{\pt}{\partial}

\newcommand{\al}{\alpha}
\newcommand{\bt}{\beta}

\newcommand{\lam}{\lambda}
\newcommand{\Lam}{\Lambda}
\newcommand{\nb}{\nabla}

\newcommand{\D}{\Delta}

\newcommand{\om}{\omega}
\newcommand{\sg}{\sigma}
\newcommand{\kp}{\kappa}

\makeatother
\begin{document}
%
\title{Non-locality and late-time cosmic acceleration from an ultraviolet complete theory 
\footnote{This manuscript belongs to an extension of the International Conference on Quantum Gravity 2018.} }
%
%
\author{Gaurav Narain$\,{}^1$}
\email{gaunarain@gmail.com}
\author{Tianjun Li$\,{}^{2,3}$}
\email{tli@itp.ac.cn}
\affiliation{%
${}^1$ Department of Space Science, Beihang University, Beijing 100191, China \\
${}^2$ CAS Key Laboratory of Theoretical Physics, Institute of Theoretical Physics, 
Chinese Academy of Sciences, Beijing 100190, China \\
${}^3$ School of Physical Sciences, University of Chinese Academy of Sciences, 
No. 19A Yuquan Road, Beijing 100049, China
}

%
%
\begin{abstract}
A local phenomenological model that reduces to a non-local gravitational 
theory giving dark energy is proposed. The non-local gravity action 
is known to fit the data as well as $\Lam$-CDM
thereby demanding a more 
fundamental local treatment. It is seen that the scale-invariant higher-derivative 
scalar-tensor theory of gravity, which is known to be ultraviolet perturbative renormalizable to 
all loops and where ghosts become innocuous, generates non-locality at low 
energies. The local action comprises of two real scalar fields coupled 
non-minimally with the higher-derivative gravity action. 
When one of the scalar acquiring the Vacuum Expectation Value (VEV)
induces Einstein--Hilbert gravity, generates mass for fields, and
gets decoupled from system, it leaves behind a residual theory which 
in turn leads to a non-local gravity generating dark energy effects.    
\end{abstract}

%

\maketitle
%
%

\section{Introduction}
\label{intro}

Dark Energy (DE) is a puzzling problem observed in the Universe 
at cosmological scales, where the Universe undergoes accelerated 
expansion \cite{Riess:1998cb,Perlmutter:1998np}. 
Numerous efforts to understand the fundamental nature of it has been considered: 
quintessence \cite{Wetterich:1987fm,Frieman:1995pm,Zlatev:1998tr,Ferreira:1997au}, 
$\Lam$-CDM, K-essence \cite{Garriga:1999vw,ArmendarizPicon:1999rj,ArmendarizPicon:2000dh}, 
but so far $\Lam$-CDM is a winning candidate fitting the data. 

It is reasonably well understood that a slowly rolling scalar give rise to 
negative pressure generating dark energy effects, which is the case in 
quintessence model \cite{Wetterich:1987fm,Frieman:1995pm,Zlatev:1998tr,Ferreira:1997au}
and k-essence model \cite{Garriga:1999vw,ArmendarizPicon:1999rj,ArmendarizPicon:2000dh}. 
Beside the existence of other possibilities, an interesting proposal 
involving non-locality has been proposed recently 
\cite{Maggiore:2014sia,Cusin2016nzi,Maggiore2016gpx}, where 
it was noticed that a particular kind of non-locality generates 
late-time accelerated expansion in the Universe fitting the dark-energy 
data as well as $\Lam$-CDM \cite{Dirian:2014bma,Dirian:2016puz}. 
However, a fundamental local explanation for it is currently 
lacking. Although, such non-localities can possibly arises 
as a consequence of quantum effects leading to energy dependence 
in coupling parameters ($g(\mu) \to g(-\Box)$, 
where $\Box$ is the square of covariant
derivative while $g$ is some coupling of the theory), where 
the infrared behaviour of the running couplings leads to 
non-local modification of gravity \cite{Maggiore:2015rma,Maggiore:2016fbn}, 
it was only recently \cite{Narain:2017twx} a model has been proposed where 
non-locality arises via decoupling of fields starting from a 
fundamental theory. 

The scale-invariant higher-derivative induced gravity model coupled 
with scalars, undergoes symmetry breaking via quantum corrections 
\cite{Narain2016}, thereby inducing mass for the scalar leading to its 
decoupling \cite{Appelquist:1974tg, Gorbar:2002pw}.
In a model comprising of more than one scalar, 
if one of the scalar gets decoupled, a phase of accelerated 
expansion is achieved at late times arising due to the 
generation of non-locality following the integration of 
residual scalars \cite{Cusin2016nzi, Dirian:2016puz, Maggiore2016gpx}. 

The outline of paper is as follows. In section \ref{nonlocal} and  \ref{induced} 
a brief review of the non-local gravity and induced gravity respectively is given. 
Section \ref{model} covers on the non-local model of gravity, while 
\ref{conc} presents conclusions.

\section{Non-local Gravity}
\label{nonlocal}

In non-local gravity, the purpose is to implement a modification in the behaviour of gravity 
at very large distances or late times of Universe. 
First proposal came in \cite{ArkaniHamed:2002fu} where 
the General Relativity (GR) equations were phenomenologically modified to 
\beq
\label{eq:GRmodify}
\left(1 - \mu^2 \Box^{-1} \right) G_\mn = 8 \pi G T_\mn \, ,
\eeq
where $\mu^2$ is the scale at which non-locality 
enters, $G_\mn$ is the Einstein tensor for the metric, $G$ is the 
gravitational coupling constant, and $T_\mn$ is the energy-momentum 
tensor. Lack of commutativity of $\nb_\mu$ 
with $\Box^{-1}$ on curved space-time implies non-conservation of 
energy-momentum tensor, indicating a much needed modification of above 
\cite{Jaccard:2013gla}. However this 
leads to unstable cosmic evolution \cite{Maggiore:2013mea,Foffa:2013vma}.
A successful non-local model, free of instabilities 
was given in \cite{Maggiore:2013mea}
\beq
\label{eq:GRmodifyTT}
G_\mn - \frac{\mu^2}{3} \left(g_\mn \Box^{-1} R \right)^T  = 8 \pi G T_\mn \, .
\eeq
This model being devoid of Veltman-Zakharov 
discontinuity, makes a smooth transition to GR when 
$\mu^2 \to 0$. Beside having a stable cosmic evolution at various eras, 
the non-locality at late-times is seen to gives rise to dark energy 
\cite{Maggiore:2013mea,Foffa:2013vma}.
Furthermore, it produces a good behaviour of cosmological perturbations in 
scalar \cite{Dirian:2014ara} and tensor sector 
\cite{Dirian:2016puz}, 
agrees very well with the CMB, supernova, Baryon 
Acoustic Oscillations (BAO) and structure formation data
\cite{Dirian:2014ara,Barreira:2014kra}. 
This model named $RT$-model (where $R$ stands for 
Ricci Scalar while $T$ refers to transverse part) is seen to be
statistically indistinguishable from $\Lam$-CDM \cite{Dirian:2014bma,Dirian:2016puz}.
This model interestingly is closely related to following non-local action,
\beq
\label{eq:nonLocalGR}
S_{\rm NL} = \frac{M_P^2}{2} \int \, {\rm d}^4x \sqrt{-g} 
\biggl[
R - \frac{\mu^2}{6} R \Box^{-2} R
\biggr] \, ,
\eeq
where $M_P$ is the reduced Planck mass, which generates 
equation-of-motions (EOM) whose linearisation around flat space 
matches the ones obtained from Eq. (\ref{eq:GRmodifyTT}), lacking 
a match at non-linear level. This model named $RR$-model works well 
at background level \cite{Maggiore:2014sia}, matching with data
\cite{Dirian:2014ara,Dirian:2016puz} 
(although it does not fit as well as $\Lam$-CDM). Exploiting the 
arbitrariness associated in describing the non-local model at the 
level of action, it is seen that the following non-linear extension 
\beq
\label{eq:nonLocalconfGR}
S_{\rm NL} = \frac{M_P^2}{2} \int \, {\rm d}^4x \sqrt{-g} 
\biggl[
R - \frac{\mu^2}{6}  R \left(-\Box + \xi R \right)^{-2} R
\biggr] \, ,
\eeq
(where $\xi$ is a dimensionless parameter) does 
a very good job in matching the data, and they get 
inspired from the realisation that the $RT$-model
is a non-linear extension of $RR$-model \cite{Cusin2016nzi}. 
This model has been studied extensively and can 
fit the DE data nicely (statistically as well as 
$\Lam$-CDM). 

\section{Induced Gravity Model}
\label{induced}

The induced gravity model comprises a scalar field non-minimally coupled
with higher-derivative gravity, whose action is of fourth order. 
The corresponding quantum theory is known to be 
renormalizable to all loops \cite{Stelle77}, 
and was recently shown to be unitary \cite{NarainA1, NarainA2} 
(see also references therein). It then offers a sufficiently  
simple quantum field theory of gravity which can be 
used to investigate physics at ultra-high energies. 

In this scale-invariant model the scalar acquires a VEV giving rise to 
gravitational coupling and masses for the fields. Lack of any dimensionful 
parameter in the scale-invariant theory makes it perturbatively renormalizable 
to all loops \cite{Fradkin1982}.
Some of the first studies of such systems were done in 
\cite{Julve1978,Fradkin1982,Barth1983, Avramidi1985}, 
where the behaviour of parameters under radiative corrections were done. 

Recently this topic has gained renewed interest 
\cite{Strumia1,Einhorn2014,Jones1,Jones2,Jones2015}.
Here a scale is generated dynamically, starting from a 
scale-invariant system by transforming the system from 
Jordon to Einstein frame where the potential acquires 
a VEV resulting in scale-generation. However, unitarity 
isn't explored
\footnote{
In \cite{Salvio2015} a quantum mechanical treatment 
of higher-derivative theories is attempted, whose suitable 
generalisation is expected to offer a treatment of 
ghosts in higher-derivative gravity.} 
It is seen in \cite{Narain2016} that the Jordan frame theory 
under quantum corrections induce ghost mass which is always 
above energy scale, thereby making it innocuous. 
The renormalizable and UV well defined scale-invariant action that one  
considers here is 
\beq
\label{eq:Act}
S_{\rm GR} = \int {\rm d}^4x \sqrt{-g} \biggl[ \frac{1}{16\pi} \biggl\{
- \frac{1}{f^2}\left(R_\mn R^\mn - \frac{1}{3}R^2 \right) 
+ \frac{\om R^2}{6f^2} \biggr\}
+ \frac{1}{2} \pt_\mu \phi \pt^\mu \phi - \frac{\lam}{4} \phi^4 - \frac{\xi}{2} R \phi^2 
\biggr] \, ,
\eeq
where the coupling $f^2$, $\om$, $\lam$ and $\xi$ are all 
dimensionless, while the curvature and covariant-derivative
depend on metric $g_\mn$. Decomposing the metric $g_\mn = \bg_\mn + h_\mn$
(where $\bg_\mn$ is background and $h_\mn$ is fluctuation), leads to 
propagator of the fluctuation field $h_\mn$ and its various couplings.
On flat space-time the propagator of quantum metric fluctuations in momentum 
space (for the Landau gauge condition $\pt^\mu h_\mn=0$) is given by,
\beq
\label{eq:GR_prop}
D^{\mn \rho\sg} = (\D_G^{-1})^{\mn\rho\sg} 
=  (16 \pi) \frac{f^2}{q^4} \left(
- 2 P_2^{\mn \rho\sg}
+ \frac{1}{\omega} P_s^{\mn\rho\sg} \right) , 
\eeq
where $(P_2)_{\mu\nu}{}^{\alpha \beta}
 = \left[ T_{\mu}{}^{\alpha} T_{\nu}{}^{\beta} + 
T_{\mu}{}^{\beta}T_{\nu}{}^{\alpha} \right]/2 - (1/(d-1))T_{\mu\nu}T^{\alpha\beta}$
and $(P_s)_{\mu\nu}{}^{\alpha \beta} 
= (1/(d-1)) T_{\mu\nu} \, T^{\alpha \beta}$ are spin 
projectors, while $L_{\mu\nu} = q_{\mu} \, q_{\nu}/q^2$
and $T_{\mu\nu} = \eta_{\mn} - q_{\mu} \, q_{\nu}/q^2$.
The coupling signs are chosen such that tachyons don't 
get generated after symmetry breaking 
\cite{Narain2016}, which implies $f^2>0$, $\om>0$, $\lam>0$ and 
$\xi>0$ (for the details on this choice of signs see \cite{Narain2016, Narain:2017tvp}). 
VEV generated via quantum corrections leads to generation of mass 
$m_s^2 = 3/2 \lam \kp^2$ and an effective Newton's constant 
$G^{-1} = 8 \pi \xi \kp^2$ with the right sign.
The graviton propagator after the symmetry breaking is following \cite{Narain2016} 
\beq
\label{eq:GRprop_sb}
D^{\mn,\al\bt} = 16 \pi G\cdot \Biggl[
\frac{(2P_2 - P_s)^{\mu\nu, \alpha\beta}}{q^2 + i \, \epsilon}
+ \frac{(P_s)^{\mu\nu, \alpha\beta}}{q^2 - M^2/\omega + i \epsilon}
- \frac{2 \, (P_2)^{\mu\nu, \alpha\beta}}{q^2 - M^2+ i \epsilon}
\Biggr] \, ,
\eeq
where the masses $M^2= 8 \pi f^2 \cdot \xi \kp^2$ 
and $M^2/\om = 8 \pi f^2 \xi \kp^2/\om$. It 
clearly shows the instability caused if $\om<0$
due to generation of tachyons \cite{NarainA1,NarainA2,Narain2016,Narain:2017tvp}.
The last term of the propagator in the broken phase (\ref{eq:GRprop_sb}) 
has a wrong sign leading to trouble with unitarity, which
is a consequence of higher-derivatives. If the generated mass 
of this ghost is such thats its mass remains always above the 
energy scale, then it becomes innocuous. Pure 
higher-derivative gravity without matter \cite{NarainA1,NarainA2}
indeed witnesses such a phenomena making the ghost innocuous 
for a large domain of coupling parameter space. This is repeated 
for the case of induced gravity \cite{Narain2016}. However, in the 
case of induced gravity it is furthermore seen that the 
scalar which acquires a VEV, gets decoupled from the system 
in the same manner as for the higher-derivative ghost.
Such a decoupling implies natural consequences in cosmology.

\section{Local to non-local}
\label{model}

The kind of non-local action stated in Eq. (\ref{eq:nonLocalconfGR}) 
is seen to naturally arise in a generalised induced gravity scenario \cite{Narain:2017twx}. 
In induced gravity scenario \cite{Narain:2017twx}, the scalar responsible 
generating VEV, mass scale, Newton's constant, gets decoupled from the system 
when the higher-derivative ghosts are evicted.
Therefore such decoupling can be exploited to construct models 
giving rise to non-localities in deep infrared. A 
two-scalar field model coupled non-minimally 
to gravity is considered
\bea
\label{eq:coupleAct}
S = \int \, {\rm d}^4x \sqrt{-g} \biggl[
&&
\frac{1}{16\pi} \biggl\{
- \frac{1}{f^2}\left(R_\mn R^\mn - \frac{1}{3}R^2 \right) 
+ \frac{\om R^2}{6f^2}  \biggr\}
+ \frac{R}{2} \Phi^T \xi \Phi 
\notag \\
&&
+ \frac{1}{2} \Phi^T (-\Box) \Phi 
- V(\Phi^T \Phi) \biggr]\, ,
\eea
where $\Phi = \{\phi, \chi \}$ is a two real scalar field doublet. It is 
an extension of model stated in Eq. (\ref{eq:Act}), where the signs of parameters 
are taken as before to ensure stability and avoid tachyons. The dimensionless 
parameter $\xi$ is a matrix whose entries are given by $\xi=\{\{\xi_1, \xi_{12}\},
\{\xi_{21}, \xi_2\}\}$. This is a renormalizable scale-invariant action 
\cite{Stelle77,Fradkin1982}, where the addition of an extra scalar 
doesn't hinders the well-defined ultraviolet behaviour of theory
and unitarity \cite{Narain2016}. 
The $\phi^4$ form of potential is considered as it obeys 
scale symmetry and renormalizability, this means 
$V(\Phi^T \Phi) = \frac{1}{4} (\Phi^T \cdot \lam \cdot \Phi)^2$,
where the coupling matrix $\lam = \{ \{\lam_1, \lam_{12} \}, \{ \lam_{21}, \lam_2 \} \}$ 
consist of only dimensionless entries. Here one of the 
scalar $\phi$ acquires VEV following Coleman-Weinberg procedure
as in \cite{Narain2016}, which leads to generation of masses 
and effective Newton's constant. The fluctuations around VEV 
denoted by $\varphi$ gets decoupled from the system eventually 
which is noticed from renormalisation group evolution \cite{Narain2016}. 
The fluctuation field $\varphi$ couples with the other scalar field $\chi$
and its action is given by,
\bea
\label{eq:symbreakAct}
&& 
S = \frac{1}{2} \int \, {\rm d}^4x \sqrt{-g} \biggl[
\kp^2 \xi_1 R + 2 \kp \xi_1 \varphi R + \varphi (-\Box + \xi_1 R) \varphi 
+ \kp (\xi_{12} + \xi_{21})R \chi
\notag \\
&&
+ \varphi(-2\Box + \xi_{12} R + \xi_{21} R )\chi
+ \chi (-\Box + \xi_2 R) \chi
-\frac{1}{2} \bigl\{
\lam_1 \kp^2 + 2 \lam_1 \kp \varphi + (\lam_{12} + \lam_{21}) \kp \chi
\notag \\
&&
+ \lam_1 \varphi^2 + (\lam_{12} + \lam_{21}) \varphi \chi 
+ \lam_2 \chi^2 
\bigr\}^2
\biggr] \, .
\eea
In special scenario when $(\lam_{12} + \lam_{21})=0$,
many terms disappear leading to a simplified form. 
In this special case mixing disappears with the induced mass for 
the fields $\varphi$ and $\chi$ given by 
$m_1^2 = 3 \lam_1^2 \kp^2$ and $m_2^2 = \lam_1 \lam_2 \kp^2$ respectively. 
The interaction piece in this special case is
\beq
\label{eq:intsimp}
I = -\frac{1}{4} \lam_1^2 \kp^4 - \lam_1^2 \kp^3 \varphi
- \lam_1 \kp \varphi \bigl\{
\lam_1 \varphi^2 + \lam_2 \chi^2 
\bigr\}
- \frac{1}{4} \bigl \{
\lam_1^2 \varphi^4 + 2 \lam_1 \lam_2 \varphi^2 \chi^2 
+ \lam_2^2 \chi^4 
\bigr\} \, .
\eeq
Following \cite{Narain:2017twx}, the dynamical evolution 
equations for fields $\varphi$ (fluctuation) and $\chi$ is obtained 
by varying the full residual action.
Although these are coupled non-linear differential equations, they 
achieve simplification with the decoupling of scalar $\varphi$ and 
for small interaction strengths (which is the case in cosmological 
scenarios). In scale-invariant higher-derivative gravity 
Eq. (\ref{eq:Act}), this phenomena naturally occurs when 
scale-symmetry gets broken. In this case the induced 
mass of the scalar is always above the energy scale leading to its 
decoupling from the system \cite{Narain2016}.
This implies $m_1^2/E^2>1$
($E$ is the running energy), which physically means that 
the particle doesn't go on-shell. In deep infrared it is seen that 
this ratio $m_1^2/E^2 \gg 1$ gets translated to 
$m_1^2 \gg \Box$. This offers simplification in our system of equations
\bea
\label{eq:largeM1_1}
&&
\varphi = -\frac{1}{3}\kappa - \frac{1}{m_1^2} \left(
\Box \chi - \frac{(\xi_{12} + \xi_{21})R \chi}{2} 
\right) \, , \\
&&
\label{eq:largeM1_2}
\left(-\Box + \xi_2 R - \frac{4}{9} m_2^2 \right) \chi 
+ \frac{(\xi_{12} + \xi_{21})R}{3} \kp = 0 \, ,
\eea
where we have ignored non-linear interactions which are higher-order.
The second equation can be easily solved after inverting the operator.
This then appears as a constraint in the system after decoupling of the 
$\varphi$ has occurred. 
Plugging back the solution for $\varphi$ from Eq. (\ref{eq:largeM1_1})
in the action, generates leading and sub-leading terms. The sub-leading part 
(which is of {\cal O}($1/m_1^2$))
can be safely ignored under decoupling. 
To leading order the action is given by,
\beq
\label{eq:residual}
S = \int \, {\rm d}^4x 
\sqrt{-g} \biggl[
- \frac{4}{81} m_1^2 \kp^2 + \frac{2 \xi_1}{9} \kp^2 R 
+ \frac{\xi_{12} + \xi_{21}}{3} \kp R \chi 
+ \frac{1}{2} (\pt \chi)^2 + \frac{\xi_2}{2} R \chi^2
- \frac{2}{9} m_2^2 \chi^2
-\frac{1}{4} \lam_2^2 \chi^4 
\biggr]\, .
\eeq
Decoupling results in a residual action for field $\chi$, the matter field 
$\chi$ couples non-minimally with the with the background space-time. 
Although symmetry breaking 
generates a large cosmological constant, but its effects gets 
shielded under the vanishing of gravitational coupling in infrared,
which happens in induced gravity coupled with 
higher-derivatives (see Eq. (73), Fig. 7, Fig. 11 and Fig. 13 of \cite{Narain2016}). 
Even in pure higher-derivative gravity 
\cite{NarainA1,NarainA2} (and in case of 
gravity coupled with gauge fields \cite{Narain:2012te,Narain:2013eea})
such vanishing of gravitational coupling is seen. 
Moreover, the field $\chi$ slow rolls when 
coupling $\lam_2$ and mass $m_2^2$ is small, leading to ignoring of the kinetic 
term of the field $\chi$. 
The field $\chi$ no longer has dynamics in this slow-roll regime. Its coupling 
with the background picture drives the dynamics of space-time. 
In this if the following approximation holds
\beq
\label{eq:finetune}
\frac{2\xi_1}{3(\xi_{12} + \xi_{21})} \gg 
\frac{\chi}{\kp} \gg \frac{2(\xi_{12} + \xi_{21})}{3\xi_2} \, ,
\eeq
then the linear $\chi$ term contribution gets suppressed by contributions from quadratic 
and quartic piece in $\chi$, and the induced Einstein-Hilbert piece \cite{Narain:2017twx}. 
This leave us with a reduced action for field $\chi$.
\beq
\label{eq:residualact1}
S = \int {\rm d}^4x \sqrt{-g} \biggl[
\frac{2 \xi_1}{9} \kp^2 R + \frac{\xi_2}{2} R \chi^2
- \frac{2}{9} m_2^2 \chi^2
-\frac{1}{4} \lam_2^2 \chi^4 
\biggr] \, .
\eeq
This action however comes with constraint Eq. (\ref{eq:largeM1_2}).
Plugging back this constraint in this residual action leads to
a non-local version of the theory
\bea
\label{eq:SNLred}
&&
S_{\rm NL} =  \int {\rm d}^4x \sqrt{-g} \biggl[
\frac{2 \xi_1}{9} \kp^2 R 
- \frac{2 m_2^2 \kp^2 (\xi_{12} + \xi_{21})^2}{81} 
R \left(-\Box + \xi_2 R - \frac{4}{9}m_2^2\right)^{-2} R 
\notag \\
&&
+ \frac{\xi_2 \kp^2 (\xi_{12} + \xi_{21})^2 R}{18} 
\biggl\{ \left(-\Box + \xi_2 R - \frac{4}{9}m_2^2\right)^{-1} R \biggr\}^2 
\notag \\
&&
\times
- \frac{\lam_2^2 \kp^4 (\xi_{12} + \xi_{21})^4}{324}  \biggl\{
\left(-\Box + \xi_2 R - \frac{4}{9}m_2^2\right)^{-1} R
\biggr\}^4
\biggr] \, . 
\eea
This is a low energy gravitational action where the first term is 
the induced gravitational term which controls the dynamics at
astronomical scales agreeing with Einstein-Hilbert gravity
while other terms are generated under decoupling and 
approximations. These are non-local gravitational interactions 
dictating the behaviour of gravitational interactions at large 
cosmological scales. It can be understood as a 
heuristic derivation of the non-local action that has been studied extensively in 
\cite{Cusin2016nzi,Maggiore2016gpx,Dirian:2016puz}
and is seen to agree with the various cosmological data 
as good as $\Lam$-CDM. 
Defining the induced Newton's constant as 
$9/(32\pi G)^{-1} = \xi_1 \kp^2$, and pulling
out the factor of Planck's mass from the action 
allows to make comparison with the existing models
in \cite{Cusin2016nzi,Maggiore2016gpx,Dirian:2016puz}. 
This will become 
\bea
\label{eq:NLactMM}
&&
S_{\rm NL} =  \frac{M_p^2}{2} \int {\rm d}^4x \sqrt{-g} \biggl[
R - \frac{\mu^2}{6} R \left(-\Box + \xi_2 R - \frac{4}{9}m_2^2\right)^{-2} R 
\notag \\
&&
+ \rho_1^2 R \biggl\{\left(-\Box + \xi_2 R - \frac{4}{9}m_2^2\right)^{-1} R \biggr\}^2
- \rho_2^2 \biggl\{\left(-\Box + \xi_2 R - \frac{4}{9}m_2^2\right)^{-1} R\biggr\}^4
\biggr] \, ,
\eea
where $M_P^2 = (4\xi_1 \kp^2)/9$ is the reduced Planck's mass and
\beq
\label{rho1rho2}
\mu^2 = \frac{2 (\xi_{12} + \xi_{21})^2 m_2^2}{3\xi_1} \, , 
\hspace{3mm}
\rho_1^2 = \frac{\xi_2 (\xi_{12} + \xi_{21})^2}{4 \xi_1} \, , 
\hspace{3mm}
\rho_2^2 = \frac{\lam_2^2 M_P^2 (\xi_{12} + \xi_{21})^4}{32 \xi_1^2} \, ,
\eeq
are the parameters defined to make connection with the 
original non-local model defined in Eq. (\ref{eq:nonLocalconfGR}). 
As seen from \cite{Cusin2016nzi,Maggiore2016gpx,Dirian:2016puz}
the mass scale $\mu$ should be quite small 
(of the ${\cal O}(H_0)$ which is ${\cal O}(10^{-32})$ eV) in order to 
explain current accelerated expansion of the Universe. 
The authors of these papers further mentions that the 
fundamental length scale that enters the physics system is $\Lam_{RR}$
and is related to $\mu^2$ via $M_P^2$ 
as $\Lam_{RR}^4 = M_P^2 \mu^2/12$. 
For our present case this implies that $\Lam_{RR}$ is related to 
the $m_2$ and $\kp$ in the following way
\beq
\label{eq:LamRR1}
\Lam_{RR}^4 = 
\frac{2}{81} (\xi_{12} + \xi_{21})^2 m_2^2 \kp^2 \, .
\eeq
This means if $\mu \sim {\cal O}(H_0)$ then 
$\Lam_{RR} \sim 10^{-3}$ eV. In well known grand unified 
theory (GUT) models, if the scale symmetry breaks around GUT scale resulting in generation of 
induced Newton's constant, then this will imply that VEV 
$\kp \sim 10^{16}$ GeV and correspondingly $\xi_1 \sim 100$. 
The mass of the scalar $\chi$ can be extracted following Eq. (\ref{eq:LamRR1}),
which is $m_2 \sim 10^{-30} (\xi_{12} + \xi_{21})^{-1}$ eV. 
Renormalizability, scale-invariance and unitarity offers 
sufficient freedom to choose parameters 
$(\xi_{12} + \xi_{21})$ allowing one to have a reasonable $m_2$. 
This generalised induced gravity model model which has a well-defined 
UV completion and ghosts are innocuous can then be seen to offer an 
interesting picture low-energy picture, where non-local interaction 
emerges in deep infrared leading to accelerated expansion at cosmological scales.

\section{Conclusion}
\label{conc}

An attempt has been made in order to understand dark energy 
which is causing the accelerated expansion of the universe. 
A local model is presented which is an ultraviolet well behaved Quantum Field Theory.
It~is a scale-invariant, perturbatively renormalizable model which is 
devoid of tachyons and where ghosts have been eradicated 
\cite{Salam1978,Julve1978, NarainA1, NarainA2, Narain2016}.
A two scalar-field model is considered which couples non-minimally with the 
scale-invariant higher-derivative gravity. One of the scalars develops a 
VEV and in turn induces low energy Einstein gravity and masses for fields. 
This scalar eventually decouples from the system as it becomes 
very massive in infrared and leaves behind a simplified system 
of residual scalar field coupled non-minimally with gravity. 
This residual local field theory model leads to non-local 
gravity action whose leading term matches the 
non-local gravity action studied extensively by 
\cite{Cusin2016nzi, Maggiore2016gpx, Barreira:2014kra, Dirian:2014bma, Dirian:2016puz}
to offer an alternative explanation for dark energy. 
This local scale-invariant system then offers a very nice 
explanation for the non-local gravity considered 
in \cite{Cusin2016nzi, Maggiore2016gpx, Barreira:2014kra, Dirian:2014bma, Dirian:2016puz}. 
The local model is not only scale invariant but also, being renormalizable and 
unitary, has a well defined UV completion.  
Furthermore, it offers a ``meaning'' to the parameter 
$\mu^2$ (or $\Lam_{RR}$) and gives a possible explanation for them.

The purpose of this work is to offer a possible local explanation for the non-local 
model considered by \cite{Cusin2016nzi, Maggiore2016gpx, 
Barreira:2014kra, Dirian:2014bma, Dirian:2016puz}. This non-local model 
has been verified against various cosmological data sets, and is 
seen to be as good as $\Lambda$-CDM. This motivated us to seek a
possible UV complete local theory which results in this particular non-local 
gravity action. Having succeeded in finding such a local theory, it is now 
important to do a more detailed study of the same. This involves, for example, 
investigating duration of the radiation/matter dominated era and other 
cosmological scenarios, which will be done in future work.

\acknowledgments{
GN and TL will like to thanks organisers Leonardo Modesto, Casimo Bambi, and 
Gianluca Calcagni for organising a wonderful conference at SUSTech. GN is very grateful 
to organisers for generously supporting his participation in conference. 
}




\begin{thebibliography}{99} 
%
%
\bibitem{Riess:1998cb}
  A.~G.~Riess {\it et al.} [Supernova Search Team],
  ``Observational evidence from supernovae for an accelerating universe and a cosmological constant,''
  Astron.\ J.\  {\bf 116} (1998) 1009
  doi:10.1086/300499
  [astro-ph/9805201].
%
%
\bibitem{Perlmutter:1998np}
  S.~Perlmutter {\it et al.} [Supernova Cosmology Project Collaboration],
  ``Measurements of Omega and Lambda from 42 high redshift supernovae,''
  Astrophys.\ J.\  {\bf 517} (1999) 565
  doi:10.1086/307221
  [astro-ph/9812133].
%
%
\bibitem{Wetterich:1987fm}
  C.~Wetterich,
  ``Cosmology and the Fate of Dilatation Symmetry,''
  Nucl.\ Phys.\ B {\bf 302} (1988) 668
  doi:10.1016/0550-3213(88)90193-9
  [arXiv:1711.03844 [hep-th]].
%
%
\bibitem{Frieman:1995pm}
  J.~A.~Frieman, C.~T.~Hill, A.~Stebbins and I.~Waga,
  ``Cosmology with ultralight pseudo Nambu-Goldstone bosons,''
  Phys.\ Rev.\ Lett.\  {\bf 75} (1995) 2077
  doi:10.1103/PhysRevLett.75.2077
  [astro-ph/9505060].
%
%
\bibitem{Zlatev:1998tr}
  I.~Zlatev, L.~M.~Wang and P.~J.~Steinhardt,
  ``Quintessence, cosmic coincidence, and the cosmological constant,''
  Phys.\ Rev.\ Lett.\  {\bf 82} (1999) 896
  doi:10.1103/PhysRevLett.82.896
  [astro-ph/9807002].
%
%
\bibitem{Ferreira:1997au}
  P.~G.~Ferreira and M.~Joyce,
  ``Structure formation with a selftuning scalar field,''
  Phys.\ Rev.\ Lett.\  {\bf 79} (1997) 4740
  doi:10.1103/PhysRevLett.79.4740
  [astro-ph/9707286].
%
%
\bibitem{Garriga:1999vw}
  J.~Garriga and V.~F.~Mukhanov,
  ``Perturbations in k-inflation,''
  Phys.\ Lett.\ B {\bf 458} (1999) 219
  doi:10.1016/S0370-2693(99)00602-4
  [hep-th/9904176].
%
%
\bibitem{ArmendarizPicon:1999rj}
  C.~Armendariz-Picon, T.~Damour and V.~F.~Mukhanov,
  ``k - inflation,''
  Phys.\ Lett.\ B {\bf 458} (1999) 209
  doi:10.1016/S0370-2693(99)00603-6
  [hep-th/9904075].
%
%
\bibitem{ArmendarizPicon:2000dh}
  C.~Armendariz-Picon, V.~F.~Mukhanov and P.~J.~Steinhardt,
  ``A Dynamical solution to the problem of a small cosmological constant and late time cosmic acceleration,''
  Phys.\ Rev.\ Lett.\  {\bf 85} (2000) 4438
  doi:10.1103/PhysRevLett.85.4438
  [astro-ph/0004134].
%
%
\bibitem{Maggiore:2014sia} 
  M.~Maggiore and M.~Mancarella,
  ``Nonlocal gravity and dark energy,''
  Phys.\ Rev.\ D {\bf 90}, no. 2, 023005 (2014)
  doi:10.1103/PhysRevD.90.023005
  [arXiv:1402.0448 [hep-th]].
%
%
\bibitem{Cusin2016nzi} 
  G.~Cusin, S.~Foffa, M.~Maggiore and M.~Mancarella,
  ``Conformal symmetry and nonlinear extensions of nonlocal gravity,''
  Phys.\ Rev.\ D {\bf 93}, no. 8, 083008 (2016)
  doi:10.1103/PhysRevD.93.083008
  [arXiv:1602.01078 [hep-th]].
%
%
\bibitem{Maggiore2016gpx} 
  M.~Maggiore,
  ``Nonlocal Infrared Modifications of Gravity. A Review,''
  Fundam.\ Theor.\ Phys.\  {\bf 187}, 221 (2017)
  doi:10.1007/978-3-319-51700-116
  [arXiv:1606.08784 [hep-th]].
%
%
\bibitem{Dirian:2014bma}
  Y.~Dirian, S.~Foffa, M.~Kunz, M.~Maggiore and V.~Pettorino,
  ``Non-local gravity and comparison with observational datasets,''
  JCAP {\bf 1504} (2015) no.04,  044
  doi:10.1088/1475-7516/2015/04/044
  [arXiv:1411.7692 [astro-ph.CO]].
%
%
\bibitem{Dirian:2016puz}
  Y.~Dirian, S.~Foffa, M.~Kunz, M.~Maggiore and V.~Pettorino,
  ``Non-local gravity and comparison with observational datasets. II. Updated results and Bayesian model comparison with $\Lambda$CDM,''
  JCAP {\bf 1605} (2016) no.05,  068
  doi:10.1088/1475-7516/2016/05/068
  [arXiv:1602.03558 [astro-ph.CO]].
%
%
\bibitem{Maggiore:2015rma}
  M.~Maggiore,
  ``Dark energy and dimensional transmutation in $R^2$ gravity,''
  arXiv:1506.06217 [hep-th].
%
%
\bibitem{Maggiore:2016fbn}
  M.~Maggiore,
  ``Perturbative loop corrections and nonlocal gravity,''
  Phys.\ Rev.\ D {\bf 93} (2016) no.6,  063008
  doi:10.1103/PhysRevD.93.063008
  [arXiv:1603.01515 [hep-th]].
%
%
\bibitem{Narain:2017twx}
  G.~Narain and T.~Li,
  ``Ultraviolet complete dark energy model,''
  Phys.\ Rev.\ D {\bf 97} (2018) no.8,  083523
  doi:10.1103/PhysRevD.97.083523
  [arXiv:1712.09054 [hep-th]].
%
%
\bibitem{Narain2016}
  G.~Narain,
  ``Exorcising Ghosts in Induced Gravity,''
  Eur.\ Phys.\ J.\ C {\bf 77} (2017) no.10,  683
  doi:10.1140/epjc/s10052-017-5249-z
  [arXiv:1612.04930 [hep-th]].
%
%
\bibitem{Appelquist:1974tg}
  T.~Appelquist and J.~Carazzone,
  ``Infrared Singularities and Massive Fields,''
  Phys.\ Rev.\ D {\bf 11} (1975) 2856.
  doi:10.1103/PhysRevD.11.2856
%
%
\bibitem{Gorbar:2002pw}
  E.~V.~Gorbar and I.~L.~Shapiro,
  ``Renormalization group and decoupling in curved space,''
  JHEP {\bf 0302} (2003) 021
  doi:10.1088/1126-6708/2003/02/021
  [hep-ph/0210388].
%
%
\bibitem{ArkaniHamed:2002fu}
  N.~Arkani-Hamed, S.~Dimopoulos, G.~Dvali and G.~Gabadadze,
  ``Nonlocal modification of gravity and the cosmological constant problem,''
  hep-th/0209227.
%
%
\bibitem{Jaccard:2013gla} 
  M.~Jaccard, M.~Maggiore and E.~Mitsou,
  ``Nonlocal theory of massive gravity,''
  Phys.\ Rev.\ D {\bf 88}, no. 4, 044033 (2013)
  doi:10.1103/PhysRevD.88.044033
  [arXiv:1305.3034 [hep-th]].
%
%
\bibitem{Maggiore:2013mea}
  M.~Maggiore,
  ``Phantom dark energy from nonlocal infrared modifications of general relativity,''
  Phys.\ Rev.\ D {\bf 89} (2014) no.4,  043008
  doi:10.1103/PhysRevD.89.043008
  [arXiv:1307.3898 [hep-th]].
%
%
\bibitem{Foffa:2013vma}
  S.~Foffa, M.~Maggiore and E.~Mitsou,
  ``Cosmological dynamics and dark energy from nonlocal infrared modifications of gravity,''
  Int.\ J.\ Mod.\ Phys.\ A {\bf 29} (2014) 1450116
  doi:10.1142/S0217751X14501164
  [arXiv:1311.3435 [hep-th]].
%
%
\bibitem{Dirian:2014ara}
  Y.~Dirian, S.~Foffa, N.~Khosravi, M.~Kunz and M.~Maggiore,
  ``Cosmological perturbations and structure formation in nonlocal infrared modifications of general relativity,''
  JCAP {\bf 1406} (2014) 033
  doi:10.1088/1475-7516/2014/06/033
  [arXiv:1403.6068 [astro-ph.CO]].
%
%
\bibitem{Barreira:2014kra}
  A.~Barreira, B.~Li, W.~A.~Hellwing, C.~M.~Baugh and S.~Pascoli,
  ``Nonlinear structure formation in Nonlocal Gravity,''
  JCAP {\bf 1409} (2014) no.09,  031
  doi:10.1088/1475-7516/2014/09/031
  [arXiv:1408.1084 [astro-ph.CO]].
%
%
\bibitem{Stelle77} 
  K.~S.~Stelle,
  ``Renormalization of Higher Derivative Quantum Gravity,''
  Phys.\ Rev.\ D {\bf 16}, 953 (1977).
  doi:10.1103/PhysRevD.16.953.
%
%
\bibitem{Salam1978} 
  A.~Salam and J.~A.~Strathdee,
  ``Remarks on High-energy Stability and Renormalizability of Gravity Theory,''
  Phys.\ Rev.\ D {\bf 18}, 4480 (1978).
  doi:10.1103/PhysRevD.18.4480
%
%
\bibitem{Julve1978} 
  J.~Julve and M.~Tonin,
  ``Quantum Gravity with Higher Derivative Terms,''
  Nuovo Cim.\ B {\bf 46}, 137 (1978).
  doi:10.1007/BF02748637
%
%
\bibitem{NarainA1} 
  G.~Narain and R.~Anishetty,
  ``Short Distance Freedom of Quantum Gravity,''
  Phys.\ Lett.\ B {\bf 711}, 128 (2012)
  doi:10.1016/j.physletb.2012.03.070
  [arXiv:1109.3981 [hep-th]].
%
%
\bibitem{NarainA2} 
  G.~Narain and R.~Anishetty,
  ``Unitary and Renormalizable Theory of Higher Derivative Gravity,''
  J.\ Phys.\ Conf.\ Ser.\  {\bf 405}, 012024 (2012)
  doi:10.1088/1742-6596/405/1/012024
  [arXiv:1210.0513 [hep-th]].
%
%
\bibitem{Fradkin1982}
  E.~S.~Fradkin and A.~A.~Tseytlin,
  ``Renormalizable asymptotically free quantum theory of gravity,''
  Nucl.\ Phys.\ B {\bf 201} (1982) 469.
  doi:10.1016/0550-3213(82)90444-8
%
%
\bibitem{Barth1983} 
  N.~H.~Barth and S.~M.~Christensen,
  ``Quantizing Fourth Order Gravity Theories. 1. The Functional Integral,''
  Phys.\ Rev.\ D {\bf 28}, 1876 (1983).
  doi:10.1103/PhysRevD.28.1876
%
%
\bibitem{Avramidi1985} 
  I.~G.~Avramidi and A.~O.~Barvinsky,
  ``Asymptotic Freedom In Higher Derivative Quantum Gravity,''
  Phys.\ Lett.\  {\bf 159B}, 269 (1985).
  doi:10.1016/0370-2693(85)90248-5
%
%
\bibitem{Strumia1}
  A.~Salvio and A.~Strumia,
  ``Agravity,''
  JHEP {\bf 1406} (2014) 080
  doi:10.1007/JHEP06(2014)080
  [arXiv:1403.4226 [hep-ph]].
%
%
\bibitem{Salvio2015}
  A.~Salvio and A.~Strumia,
  ``Quantum mechanics of 4-derivative theories,''
  Eur.\ Phys.\ J.\ C {\bf 76} (2016) no.4,  227
  doi:10.1140/epjc/s10052-016-4079-8
  [arXiv:1512.01237 [hep-th]].
%
%
\bibitem{Einhorn2014} 
  M.~B.~Einhorn and D.~R.~T.~Jones,
  ``Naturalness and Dimensional Transmutation in Classically Scale-Invariant Gravity,''
  JHEP {\bf 1503}, 047 (2015)
  doi:10.1007/JHEP03(2015)047
  [arXiv:1410.8513 [hep-th]].
%
%
\bibitem{Jones2015} 
  T.~Jones and M.~Einhorn,
  ``Quantum Gravity and Dimensional Transmutation,''
  PoS PLANCK {\bf 2015}, 061 (2015).
%
%
\bibitem{Jones1} 
  M.~B.~Einhorn and D.~R.~T.~Jones,
  ``Induced Gravity I: Real Scalar Field,''
  JHEP {\bf 1601}, 019 (2016)
  doi:10.1007/JHEP01(2016)019
  [arXiv:1511.01481 [hep-th]].
%
%
\bibitem{Jones2} 
  M.~B.~Einhorn and D.~R.~T.~Jones,
  ``Induced Gravity II: Grand Unification,''
  JHEP {\bf 1605}, 185 (2016)
  doi:10.1007/JHEP05(2016)185
  [arXiv:1602.06290 [hep-th]].
%
%
\bibitem{Narain:2017tvp}
  G.~Narain,
  ``Signs and Stability in Higher-Derivative Gravity,''
  Int.\ J.\ Mod.\ Phys.\ A {\bf 33} (2018) no.04,  1850031
  doi:10.1142/S0217751X18500318
  [arXiv:1704.05031 [hep-th]].
%
%
\bibitem{Narain:2012te} 
  G.~Narain and R.~Anishetty,
  ``Charge Renormalization due to Graviton Loops,''
  JHEP {\bf 1307}, 106 (2013)
  doi:10.1007/JHEP07(2013)106
  [arXiv:1211.5040 [hep-th]].
%
%
\bibitem{Narain:2013eea} 
  G.~Narain and R.~Anishetty,
  ``Running Couplings in Quantum Theory of Gravity Coupled with Gauge Fields,''
  JHEP {\bf 1310}, 203 (2013)
  doi:10.1007/JHEP10(2013)203
  [arXiv:1309.0473 [hep-th]].
%
%
%
\end{thebibliography}
\end{document}